\documentclass{emulateapj}

\usepackage{amssymb}

\def\ts     {\thinspace}
\def\kms    {\ts km\ts s$^{-1}$}
\def\etal   {{\rm et\ts al.}}
\def\msol   {$M_{\odot}$}
\def\lsol   {$L_{\odot}$}
\def\lprime {K\,\kms\,pc$^2$}

\def\aco    {{\rm CO}($J$=1$\to$0)}
\def\acoalt {{\rm CO} $J$=1$\to$0}

\def\cco    {{\rm CO}($J$=3$\to$2)}

\def\eco    {{\rm CO}($J$=5$\to$4)}

\def\altgco {{\rm CO} $J$=7$\to$6}

\def\naltjco {$J$=10$\to$9}

\def\gcs    {{\rm CS}($J$=7$\to$6)}

\def\dhcn    {{\rm HCN}($J$=4$\to$3)}

\def\hsio    {{\rm SiO}($J$=8$\to$7)}

\def\bhcccn    {{\rm HC$_3$N}($J$=39$\to$38)}
\def\chcccn    {{\rm HC$_3$N}($J$=63$\to$62)}
\def\dhcccn    {{\rm HC$_3$N}($J$=64$\to$63)}

\def\hlsw    {HERMES\,J105751.1+573027}

\def\pdbi     {Plateau de Bure Interferometer (PdBI)}
\def\carma    {Combined Array for Research in Millimeter-wave Astronomy (CARMA)}

\slugcomment{draft version \today, accepted for publication in the Astrophysical Journal Letters}

\shorttitle{CO Imaging of an Extremely Bright Submillimeter Galaxy at $z$$\simeq$3}
\shortauthors{Riechers et al.}

\begin{document}

\title{
Dynamical Structure of the Molecular Interstellar Medium in an
Extremely Bright, Multiply Lensed $z$$\simeq$3 Submillimeter Galaxy
Discovered with Herschel}

\author{Dominik A.\ Riechers\altaffilmark{1,28},
A.~Cooray\altaffilmark{2,1},
A.~Omont\altaffilmark{3},
R.~Neri\altaffilmark{4},
A.I.~Harris\altaffilmark{5},
A.J.~Baker\altaffilmark{6},
P.~Cox\altaffilmark{4},
D.T.~Frayer\altaffilmark{7},
J.M.~Carpenter\altaffilmark{1},
R.~Auld\altaffilmark{8},
H.~Aussel\altaffilmark{9},
A.~Beelen\altaffilmark{10},
R.~Blundell\altaffilmark{11},
J.~Bock\altaffilmark{1,12},
D.~Brisbin\altaffilmark{13},
D.~Burgarella\altaffilmark{14},
P.~Chanial\altaffilmark{9},
S.C.~Chapman\altaffilmark{15},
D.L.~Clements\altaffilmark{16},
A.~Conley\altaffilmark{17},
C.D.~Dowell\altaffilmark{12},
S.~Eales\altaffilmark{8},
D.~Farrah\altaffilmark{18},
A.~Franceschini\altaffilmark{19},
R.~Gavazzi\altaffilmark{3},
J.~Glenn\altaffilmark{17},
M.~Griffin\altaffilmark{8},
M.~Gurwell\altaffilmark{11},
R.J.~Ivison\altaffilmark{20,21},
S.~Kim\altaffilmark{2},
M.~Krips\altaffilmark{4},
A.M.J.~Mortier\altaffilmark{16},
S.J.~Oliver\altaffilmark{18},
M.J.~Page\altaffilmark{22},
A.~Papageorgiou\altaffilmark{8},
C.P.~Pearson\altaffilmark{23,24},
I.~P{\'e}rez-Fournon\altaffilmark{25,26},
M.~Pohlen\altaffilmark{8},
J.I.~Rawlings\altaffilmark{22},
G.~Raymond\altaffilmark{8},
G.~Rodighiero\altaffilmark{19},
I.G.~Roseboom\altaffilmark{18},
M.~Rowan-Robinson\altaffilmark{16},
K.S.~Scott\altaffilmark{27},
N.~Seymour\altaffilmark{22},
A.J.~Smith\altaffilmark{18},
M.~Symeonidis\altaffilmark{22},
K.E.~Tugwell\altaffilmark{22},
M.~Vaccari\altaffilmark{19},
J.D.~Vieira\altaffilmark{1},
L.~Vigroux\altaffilmark{3},
L.~Wang\altaffilmark{18},
J.~Wardlow\altaffilmark{2},
and G.~Wright\altaffilmark{20}}

\altaffiltext{1}{Astronomy Department, California Institute of
  Technology, MC 249-17, 1200 East California Boulevard, Pasadena, CA
  91125, USA; dr@caltech.edu}

\altaffiltext{2}{Department of Physics \& Astronomy, University of California, Irvine, CA 92697, USA}

\altaffiltext{3}{Institut d'Astrophysique de Paris, CNRS and Universit\'{e} Pierre et Marie Curie, 98bis Bd Arago, F-75014 Paris, France}

\altaffiltext{4}{Institut de RadioAstronomie Millim\'etrique, 300 Rue
  de la Piscine, Domaine Universitaire, F-38406 Saint Martin d'H\'eres,
  France}

\altaffiltext{5}{Department of Astronomy, University of Maryland, College Park, MD 20742-2421, USA}

\altaffiltext{6}{Department of Physics and Astronomy, Rutgers, the State University of New Jersey, 136 Frelinghuysen Road, Piscataway, NJ 08854-8019, USA}

\altaffiltext{7}{National Radio Astronomy Observatory, P.O. Box 2, Green Bank, WV 24944, USA}

\altaffiltext{8}{School of Physics and Astronomy, Cardiff University, Queens Buildings, The Parade, Cardiff CF24 3AA, UK}

\altaffiltext{9}{Laboratoire AIM-Paris-Saclay, CEA/DSM/Irfu - CNRS - Universit\'e Paris Diderot, CE-Saclay, pt courrier 131, F-91191 Gif-sur-Yvette, France}

\altaffiltext{10}{Institut d'Astrophysique Spatiale, Universit\'{e} Paris-Sud XI, b\^at 121, F-91405 Orsay Cedex, France}

\altaffiltext{11}{Harvard-Smithsonian Center for Astrophysics, Cambridge, MA 02138, USA}

\altaffiltext{12}{Jet Propulsion Laboratory, 4800 Oak Grove Drive, Pasadena, CA 91109, USA}

\altaffiltext{13}{Space Science Building, Cornell University, Ithaca, NY, 14853-6801, USA}

\altaffiltext{14}{Laboratoire d'Astrophysique de Marseille, OAMP, Universit\'e Aix-marseille, CNRS, 38 rue Fr\'ed\'eric Joliot-Curie, F-13388 Marseille cedex 13, France}

\altaffiltext{15}{Institute of Astronomy, University of Cambridge, Madingley Road, Cambridge CB3 0HA, UK}

\altaffiltext{16}{Astrophysics Group, Imperial College London, Blackett Laboratory, Prince Consort Road, London SW7 2AZ, UK}

\altaffiltext{17}{Dept. of Astrophysical and Planetary Sciences, CASA 389-UCB, University of Colorado, Boulder, CO 80309, USA}

\altaffiltext{18}{Astronomy Centre, Dept. of Physics \& Astronomy, University of Sussex, Brighton BN1 9QH, UK}

\altaffiltext{19}{Dipartimento di Astronomia, Universit\`{a} di Padova, vicolo Osservatorio, 3, I-35122 Padova, Italy}

\altaffiltext{20}{UK Astronomy Technology Centre, Royal Observatory, Blackford Hill, Edinburgh EH9 3HJ, UK}

\altaffiltext{21}{Institute for Astronomy, University of Edinburgh, Royal Observatory, Blackford Hill, Edinburgh EH9 3HJ, UK}

\altaffiltext{22}{Mullard Space Science Laboratory, University College London, Holmbury St. Mary, Dorking, Surrey RH5 6NT, UK}

\altaffiltext{23}{Space Science \& Technology Department, Rutherford Appleton Laboratory, Chilton, Didcot, Oxfordshire OX11 0QX, UK}

\altaffiltext{24}{Institute for Space Imaging Science, University of Lethbridge, Lethbridge, Alberta, T1K 3M4, Canada}

\altaffiltext{25}{Instituto de Astrof{\'\i}sica de Canarias, E-38200 La Laguna, Tenerife, Spain}

\altaffiltext{26}{Departamento de Astrof{\'\i}sica, Universidad de La Laguna, E-38205 La Laguna, Tenerife, Spain}

\altaffiltext{27}{Department of Physics and Astronomy, University of Pennsylvania, Philadelphia, PA 19104, USA}

\altaffiltext{28}{Hubble Fellow}


\begin{abstract}

We report the detection of \eco, \cco, and \aco\ emission in the
strongly lensed, {\em Herschel}/SPIRE-selected submillimeter galaxy
(SMG) \hlsw\ at $z$=2.9574$\pm$0.0001, using the Plateau de Bure
Interferometer, the Combined Array for Research in Millimeter-wave
Astronomy, and the Green Bank Telescope.  The observations spatially
resolve the molecular gas into four lensed images with a maximum
separation of $\sim$9$''$, and reveal the internal gas dynamics in
this system. We derive lensing-corrected CO line luminosities of
$L'_{\rm CO(1-0)} = (4.17 \pm 0.41)$, $L'_{\rm CO(3-2)} = (3.96 \pm
0.20)$ and $L'_{\rm CO(5-4)} = (3.45 \pm 0.20) \times
10^{10}\,(\mu_{\rm L}/10.9)^{-1}$\,\lprime, corresponding to
luminosity ratios of $r_{31}$=0.95$\pm$0.10, $r_{53}$=0.87$\pm$0.06,
and $r_{51}$=0.83$\pm$0.09.  This suggests a total molecular gas mass
of $M_{\rm gas}$=3.3$\times$10$^{10}$\,($\alpha_{\rm
CO}$/0.8)\,($\mu_{\rm L}$/10.9)$^{-1}$\,\msol.  The gas mass, gas mass
fraction, gas depletion timescale, star formation efficiency, and
specific star formation rate are typical for an SMG.  The velocity
structure of the gas reservoir suggests that the brightest two lensed
images are dynamically resolved projections of the same dust-obscured
region in the galaxy that are kinematically offset from the unresolved
fainter images.  The resolved kinematics appear consistent with the
complex velocity structure observed in major, `wet' (i.e., gas-rich)
mergers.  Major mergers are commonly observed in SMGs, and are likely
to be responsible for fueling their intense starbursts at high gas
consumption rates.  This study demonstrates the level of detail to
which galaxies in the early universe can be studied by utilizing the
increase in effective spatial resolution and sensitivity provided by
gravitational lensing.

\end{abstract}

\keywords{galaxies: active --- galaxies: starburst --- 
galaxies: formation --- galaxies: high-redshift --- cosmology: observations 
--- radio lines: galaxies}

\section{Introduction}

Great progress has been made over the past decade in understanding the
role of submillimeter-selected galaxies (SMGs; see reviews by Blain et
al.\ \citeyear{bla02}; Lagache et al.\ \citeyear{lag05}; Smail
\citeyear{sma06}) in the early evolution of massive galaxies. Studies
of the stellar populations of massive galaxies suggest that they form
the bulk of their stars at early epochs ($z$$>$2), with the more
massive galaxies building up their stellar component at earlier epochs
and on shorter timescales (see review by Renzini
\citeyear{ren06}). These findings suggest the existence of a
population of massive, gas-rich starburst galaxies at early cosmic
times -- properties which are consistent with those of SMGs. SMGs are
typically intense ($\gtrsim$500\,\msol\,yr$^{-1}$), optically
obscured, relatively short-lived ($<$100\,Myr) starbursts at high
redshift, implying rapid gas consumption timescales at high star
formation efficiencies (e.g., Smail et al.\ \citeyear{sma97}; Hughes
et al.\ \citeyear{hug98}; Greve et al.\ \citeyear{gre05}). Most SMGs
have sizes of a few kpc and are dynamically complex, exhibiting the
properties of gas-rich, major mergers (e.g., Tacconi et al.\
\citeyear{tac08}).

The physical properties of the star-forming material in these extreme
galaxies are best studied using rotational emission lines of molecular
gas tracers, most commonly CO. Molecular gas is the fuel for star
formation, and has been detected in $>$30 SMGs to date. These
typically have large gas reservoirs of $>$10$^{10}$\,\msol\ (see
review by Solomon \& Vanden Bout \citeyear{sv05}). The difficulty of
such studies is reflected in the moderate number of detections and
limited level of detail of most experiments. A promising way to
overcome these limitations is to study strongly gravitationally lensed
SMGs, as lensing provides a boost in sensitivity and spatial
resolution. However, strongly lensed SMGs are very rare, making them
difficult to find in typical submillimeter continuum surveys.

With the advent of large area ($\gg$1\,deg$^2$) submillimeter surveys
that are currently being undertaken with {\em Herschel} such as the
Multi-tiered Extragalactic Survey (HerMES; S.~Oliver et al., in
prep.),\footnote{\tt http://hermes.sussex.ac.uk/ } efficient
identification of lensed SMGs is now becoming possible. Here, we study
the molecular gas properties of the brightest source identified in
these studies.

In this Letter, we report the detection of spatially resolved \eco\
and \cco\ emission and of \aco\ emission toward the strongly lensed
SMG \hlsw\ (hereafter:\ HLSW-01; $z$=2.957), using the \pdbi, the
\carma, and the 100\,m Green Bank Telescope (GBT). HLSW-01 was
discovered in the HerMES {\em Herschel}/SPIRE Science Demonstration
Phase (SDP) data (Oliver et al.\ \citeyear{oli10}), and is the
brightest $z$$>$2 SMG currently known. Its spectral energy
distribution peaks at an (apparent) 250\,$\mu$m flux of
425$\pm$10\,mJy (Conley et al.\ \citeyear{con10}; C11). It is lensed
by a galaxy group at $z_{\rm G}$=0.60$\pm$0.04, magnifying the SMG by
a factor of $\mu_{\rm L}$=10.9$\pm$0.7 in flux density (Gavazzi et
al.\ \citeyear{gav10}; G11).  In accompanying work, we provide
additional details on the identification, molecular line excitation
(based on this work and \altgco\ to \naltjco\ observations with
CSO/Z-spec; Scott et al.\ \citeyear{sco10}; S11), spectral energy
distribution (C11) and lensing properties (G11) of HLSW-01. We use a
concordance, flat $\Lambda$CDM cosmology throughout, with
$H_0$=71\,\kms\,Mpc$^{-1}$, $\Omega_{\rm M}$=0.27, and
$\Omega_{\Lambda}$=0.73 (Spergel \etal\ \citeyear{spe03},
\citeyear{spe07}).

\begin{figure*}
\epsscale{1.15}
\plotone{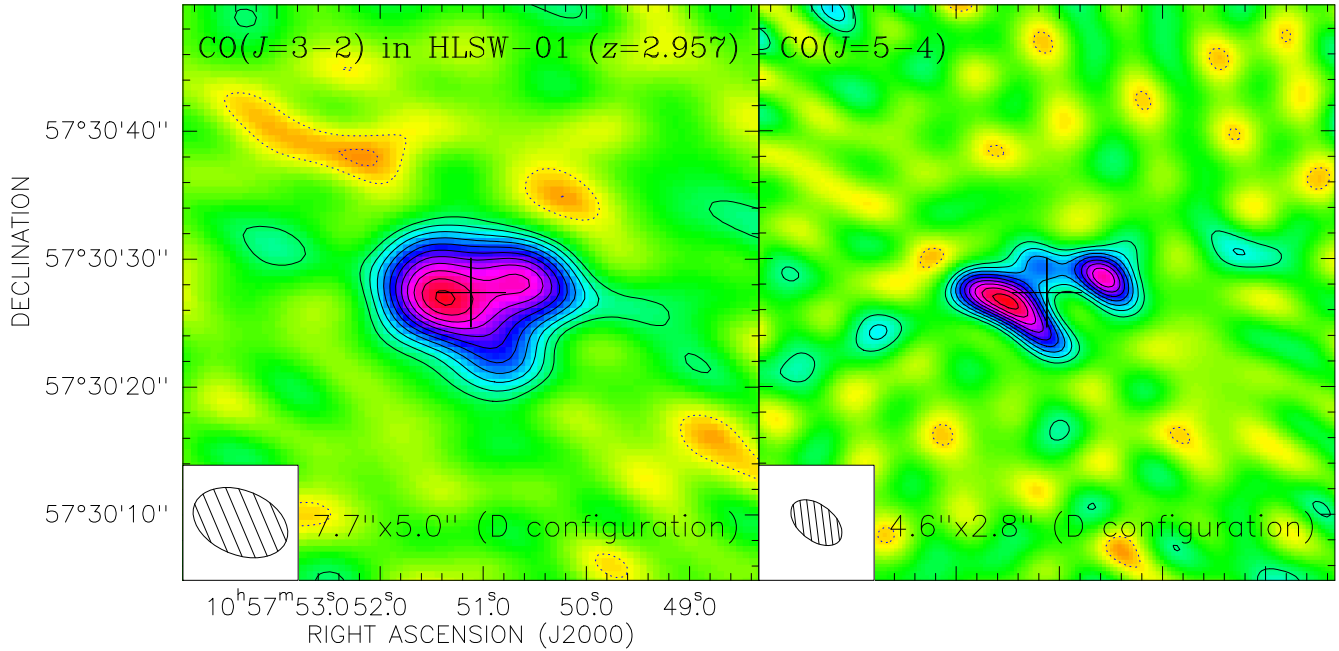}

\caption{Velocity-integrated CARMA and PdBI maps of \cco\ and \eco\ line 
emission over 777 and 515\,\kms\ toward HLSW-01. At 7.7$''$$\times$5.0$''$
and 4.6$''$$\times$2.8$''$ resolution (as indicated in the bottom
left), the emission is clearly resolved. The crosses indicate the
center of the 250\,$\mu$m emission (C11).  Contours are shown in steps
of 1$\sigma$=0.48 and 1.2\,mJy\,beam$^{-1}$, starting at
$\pm$2$\sigma$.
\label{f1}}
%
\end{figure*}

\section{Observations}

\subsection{Plateau de Bure Interferometer}

We observed the \eco\ transition line ($\nu_{\rm rest} =
576.2679305\,$GHz, redshifted to 145.633\,GHz, or 2.06\,mm) toward
HLSW-01 ($z$=2.957), using the PdBI.  We used the WideX correlator
with a total bandwidth of 3.6\,GHz ($\sim$7400\,\kms; at 1.95\,MHz
resolution) to cover the \eco\ line and the underlying 2\,mm
(rest-frame 520\,$\mu$m) continuum emission.  Observations were
carried out under good 2\,mm weather conditions for 2\,tracks in 5D
configuration on 2010 June 14 and 17. This resulted in 1.0\,hr of 6
antenna-equivalent on-source time after discarding unusable visibility
data. The nearby ($10.6^\circ$ distant) source B0954+658 (J0958+655;
0.98\,Jy) was observed every 20 minutes for pointing, amplitude and
phase calibration.  MWC\,349 (1.50\,Jy) and 3C273 (10.5\,Jy) were
observed for flux and bandpass calibration, yielding $\sim$10\%
calibration accuracy.

The IRAM GILDAS package was used for data reduction and analysis.  All
data were mapped using the CLEAN algorithm with `natural' weighting,
resulting in a synthesized beam of 4.6\,$''$$\times$2.8\,$''$.  The
final rms is 1.2\,mJy beam$^{-1}$ over 250\,MHz (corresponding to
515\,\kms), 3.4\,mJy beam$^{-1}$ over 31.25\,MHz (64\,\kms), and
6.0\,mJy beam$^{-1}$ over 10\,MHz (21\,\kms).

\subsection{CARMA}

We observed the \cco\ transition line ($\nu_{\rm rest} =
345.7959899\,$GHz, redshifted to 87.388\,GHz, or 3.43\,mm), using
CARMA.  A total bandwidth of 3.7\,GHz ($\sim$12,700\,\kms; at
5.208\,MHz resolution) was used to cover the \cco\ line and the
underlying 3.4\,mm (rest-frame 870\,$\mu$m) continuum emission.
Observations were carried out under good 3\,mm weather conditions
(typically at moderate elevations after transit) for 6\,tracks in D
configuration between 2010 September 01 and 17.  This resulted in
6.6\,hr of 15 antenna-equivalent on-source time after discarding
unusable visibility data. The nearby source J0958+655 (0.86\,Jy) was
observed every 15 minutes for pointing, amplitude and phase
calibration. Fluxes were bootstrapped relative to Mars (based on a
brightness temperature model of the planet). Several nearby
calibrators (3C84, 3C273, 3C345, J0927+390, J2015+372) were observed
for bandpass calibration, yielding $\sim$15\% calibration accuracy.

The MIRIAD package was used for data reduction and analysis.  All data
were mapped using the CLEAN algorithm with `natural' weighting;
resulting in a synthesized beam of 7.7\,$''$$\times$5.0\,$''$.  The
final rms is 0.48\,mJy beam$^{-1}$ over 226\,MHz (corresponding to
777\,\kms), 1.6\,mJy beam$^{-1}$ over 20.8\,MHz (71\,\kms), and
2.3\,mJy beam$^{-1}$ over 10.4\,MHz (36\,\kms).

\newpage

\subsection{GBT}

We observed the \aco\ transition line ($\nu_{\rm rest} =
115.2712018\,$GHz, redshifted to 29.131\,GHz, or 10.29\,mm), using the
Zpectrometer analog lag cross-correlation spectrometer (Harris et al.\
\citeyear{har07}) attached to the GBT Ka band receiver.  This yields
an instantaneous bandwidth of 10.5\,GHz ($\sim$108,000\,\kms) at a
beam size of 27$''$--19$''$ (25.6--36.1\,GHz). Observations were
carried out under good 1\,cm weather conditions on 2010 October
02. This resulted in 1.5\,hr on-source time after discarding unusable
data. Subreflector beam switching was used every 10\,s to observe the
source alternately with the receiver's two beams, with the off-source
beam monitoring the sky background in parallel. A nearby second,
fainter source was observed with the same pattern, switching between
targets with 8\,min cycles. Residual structure from optical beam
imbalance in the ``source--sky'' difference spectra of each target was
then removed by differencing both sources.  A spectral ripple with
well-defined period generated in the receiver was removed with
narrow-band Fourier filtering. This strategy yields a flat baseline
(offset from zero flux by to the difference of the two faint source
continua) without standard polynomial baseline removal.  The nearby
quasar 1143+4931 was observed every $\sim$1\,hr to monitor telescope
pointing and gain stability. Passband gains and absolute fluxes were
determined from spectra of 3C286 (2.04\,Jy at 32\,GHz; Ott et al.\
\citeyear{ott94}), yielding 15\% calibration accuracy.\footnote{The
2010 Astronomical Almanac suggests a slightly lower value of 1.95\,Jy
at 29\,GHz for 3C286.}

The lag data were processed using the Zpectrometer's data reduction
pipeline and standard recipes, yielding a frequency-calibrated
spectrum at 8\,MHz resolution (see Harris et al.\ \citeyear{har10} for
details).  The instrumental spectral response is nearly a sinc
function with an FWHM of 20\,MHz, i.e., individual 8\,MHz spectral
channels are not statistically independent. However, the line width
correction for the instrumental response for spectral lines with
intrinsic Gaussian FWHM of $>$30\,MHz ($\sim$300\,\kms ) is minor.

\begin{figure*}
\epsscale{1.15}
\plotone{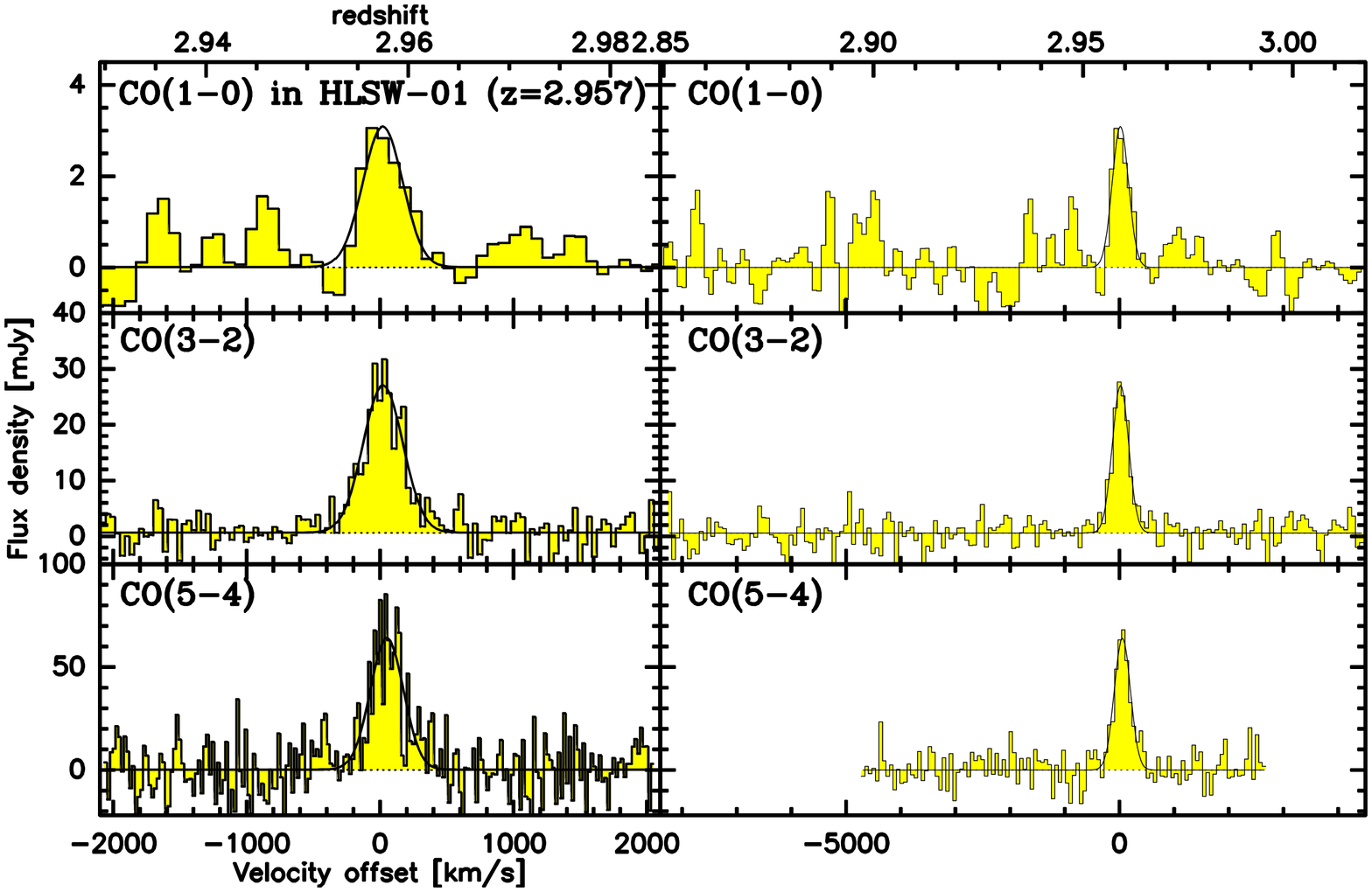}

\caption{{\em Left}:\ \aco, \cco\ and \eco\ spectra of HLSW-01 at 
82/36/21\,\kms\ (8.0/10.416/10.0\,MHz) resolution (histograms), along
with Gaussian fits to the line and continuum emission (black
curves). The velocity scale is relative to $z$=2.957. The \aco\
channels are not statistically independent on scales below 20\,MHz
(2.5\,channels). {\em Right}:\ Same, re-binned to similar velocity
resolution (82/71/64\,\kms, or 8.0/20.832/31.25\,MHz; \acoalt\
unchanged), and showing a larger velocity range.
\label{f2}}
%
\end{figure*}

\section{Results}

\subsection{Integrated Emission Line and Continuum Properties}

We have detected spatially resolved \cco\ and \eco\ (Fig.~\ref{f1})
and unresolved \aco\ line emission toward HLSW-01 at high
significance.  From Gaussian fitting to the (spatially integrated)
line profiles (Fig.~\ref{f2}), we obtain \aco, \cco, and
\eco\ line peak strengths of 3.09$\pm$0.37, 26.4$\pm$1.5 and
64.0$\pm$4.4\,mJy at a median FWHM of 348$\pm$18\,\kms\ (350$\pm$55,
350$\pm$25, and 346$\pm$29\,\kms\ for the CO $J$=1$\to$0, 3$\to$2, and
5$\to$4 lines). We marginally detect 3.43\,mm continuum emission under
the \cco\ line at 0.61$\pm$0.19\,mJy strength, and do not detect
2.06\,mm continuum emission down to a 2$\sigma$ limit of 1.4\,mJy.  By
combining all lines, we find a median redshift of
$z$=2.9574$\pm$0.0001.  The line parameters correspond to
velocity-integrated emission line strengths of 1.14$\pm$0.11,
9.74$\pm$0.49, and 23.6$\pm$1.4\,Jy\,\kms, and line luminosities of
$L'_{\rm CO(1-0)} = (4.17 \pm 0.41)$, $L'_{\rm CO(3-2)} = (3.96 \pm
0.20)$, and $L'_{\rm CO(5-4)} = (3.45 \pm 0.20)\times
10^{10}\,(\mu_{\rm L}/10.9)^{-1}\,$\lprime, respectively. This
corresponds to line brightness temperature ratios of
$r_{31}$=0.95$\pm$0.10, $r_{53}$=0.87$\pm$0.06, and
$r_{51}$=0.83$\pm$0.09. This suggests that at least the \eco\ line is
not thermalized (i.e., not in radiative equilibrium with the lower-$J$
lines; $r$$<$1; see also S11).  We do not find evidence for a
substantial low-excitation gas component in HLSW-01 (i.e., gas with
$r_{31}$$\ll$1), contrary to what is found in some other $z$$>$2 SMGs
(e.g., Carilli et al.\ \citeyear{car10}; Riechers et al.\
\citeyear{rie10}; Harris et al.\ \citeyear{har10}; Ivison et al.\
\citeyear{ivi10a}). We estimate the total molecular gas mass assuming
a conversion factor of $\alpha_{\rm CO}$=0.8\,\msol\,(\lprime )$^{-1}$
from $L'_{\rm CO(1-0)}$ to $M_{\rm gas}$, as appropriate for
ultra-luminous infrared galaxies (ULIRGs; Downes \& Solomon
\citeyear{ds98}) and commonly used for SMGs. This yields $M_{\rm
gas}$=3.3$\times$10$^{10}$\,($\alpha_{\rm CO}$/0.8)\,($\mu_{\rm
L}$/10.9)$^{-1}$\,\msol.

\subsection{Limits on Dense Gas Tracers}

Several lines of dense molecular gas tracers fall within the spectral
range covered by the \cco\ and \eco\ observations.  We obtain
3$\sigma$ upper limits of $<$1.6\,Jy\,\kms\ on the integrated
\dhcn, \gcs, \hsio, and \bhcccn\ emission line strengths, and 
$<$3.2\,Jy\,\kms\ for \chcccn\ and \dhcccn\ (averaged over 348\,\kms
).  This corresponds to line luminosity limits of $<$0.61, $<$0.66,
$<$0.64, $<$0.61, $<$0.48, and $<$0.46 $\times 10^{10}\,(\mu_{\rm
L}/10.9)^{-1}\,$\lprime\ for the HCN, CS, SiO, \bhcccn, \chcccn, and
\dhcccn\ lines, respectively. This also corresponds to line brightness
temperature ratio limits of $<$0.15, $<$0.16, $<$0.15, $<$0.15,
$<$0.11, and $<$0.11 relative to \aco. Nearby starburst galaxies and
ULIRGs like NGC\,253, NGC\,6240, and Arp\,220 typically have HCN
$J$=4$\to$3/$J$=1$\to$0 line ratios of 0.45--0.8 (Knudsen et al.\
\citeyear{knu07}; Greve et al.\ \citeyear{gre09}). If this ratio is
similar in HLSW-01, it suggests a HCN/CO ratio of $<$0.18--0.33 in the
respective $J$=1$\to$0 transitions, consistent with what is found in
nearby ULIRGs and other high-$z$ gas-rich galaxies (median ratios of
0.17; Gao \& Solomon \citeyear{gs04a}, \citeyear{gs04b}; Gao et al.\
\citeyear{gao07}; Riechers et al.\ \citeyear{rie07b}).

\begin{figure*}
\epsscale{1.15}
\plotone{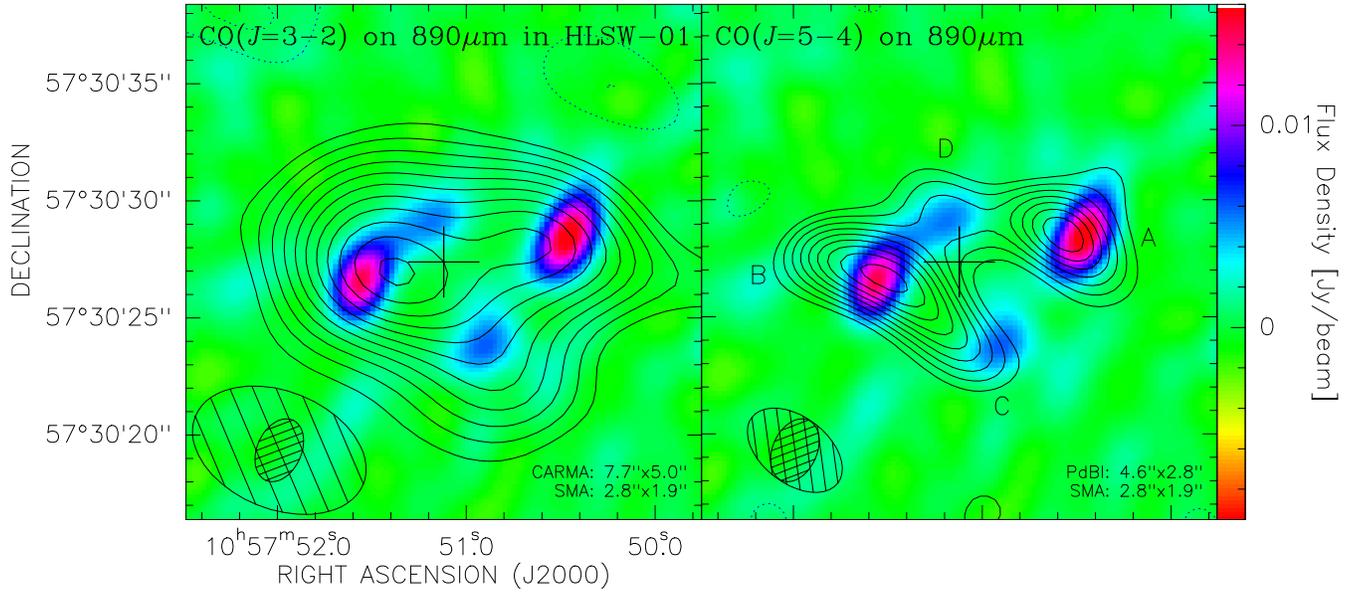}

\caption{\cco\ and \eco\ line emission contours as shown in 
Fig.~\ref{f1}, but overlayed on 890\,$\mu$m continuum emission
detected with the Submillimeter Array (SMA; color scale;
2.8$''$$\times$1.9$''$ resolution, as indicated in the bottom left;
C11). The crosses indicate the same position as in Fig.~\ref{f1}. `A'
to `D' label the four lensed images of the source ordered by
brightness of the 890\,$\mu$m continuum emission. CO emission is
detected toward all four images.
\label{f3}}
%
\end{figure*}

\subsection{Morphology of the Molecular Gas Reservoir}

We spatially resolve the (lensed) molecular gas reservoir in HLSW-01
in the \cco\ and \eco\ emission lines. The characteristic effective
resolution provided by the lensing effect is $\sim$2.2$''$ (i.e.,
regions of $\sim$8.5\,kpc radius) in \eco, but substantially higher in
the brighter images and close to critical lines (G11). Overlays of the
CO emission with 890\,$\mu$m (rest-frame 225\,$\mu$m) continuum
emission are shown in Fig.~\ref{f3}.  The rest-frame far-infrared
(FIR) continuum emission dominantly traces the regions of most active
star formation in this system (star formation rate:\
2460$\pm$160\,($\mu_{\rm L}$/10.9)$^{-1}$\,\msol\,yr$^{-1}$; C11). The
\eco\ emission shows discernible peaks at the positions of lens images
A, B, and D, with an extension of the emission associated with image B
toward image C. Accounting for beam convolution, there is no evidence
for differences in the spatial distribution and/or line shapes of the
\cco\ and \eco\ emission, indicating that both lines trace the same
material.  The small offsets between the CO and FIR peaks (typically
$<$10\% of the relative beam sizes) are a combination of beam
convolution (given the different beam sizes and orientations) and
velocity-averaged dynamical structure in the integrated molecular line
maps. The molecular gas reservoir may extend beyond the FIR continuum
peaks, but maps at higher spatial resolution and sensitivity are
required to investigate this in more detail. The overall striking
match between the gas and (FIR) dust emission suggests that, indeed,
the bulk of the FIR light is associated with the gas-rich,
star-forming regions.

\begin{figure}
\vspace{-7.25mm}

\epsscale{1.45}
\plotone{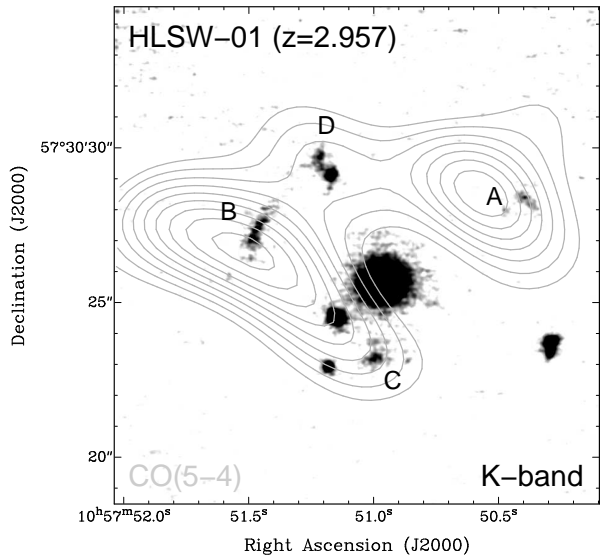}

\caption{\eco\ line emission contours as shown in 
Fig.~\ref{f1}, but overlayed on 2.2\,$\mu$m continuum emission (gray
scale; smoothed with a 0.1$''$ Gaussian kernel; C11). `A' to `D' label
the four lensed images as in Fig.~\ref{f3}. The CO emission is
associated with the four rest-frame optical lensed images, but the
peaks of the emission are optically faint.
\label{f4}}
%
\end{figure}

An overlay of the \eco\ emission with 2.2\,$\mu$m (rest-frame 560\,nm;
smoothed with a 0.1$''$ Gaussian kernel) continuum emission is shown
in Fig.~\ref{f4}. The CO emission is clearly associated with the faint
optical lensed images of the SMG, which appear to consist of multiple
clumps.  However, the brightest CO peaks are slightly offset from the
peaks of the rest-frame optical emission. This offset translates to
$\sim$2.4\,kpc in the source plane after correcting for gravitational
magnification (G11).  This may indicate that the most gas-rich regions
are optically obscured and/or the presence of multiple galaxy
components, as commonly observed in SMGs. Also, the FIR-brightest lens
image A is the optically faintest image of the galaxy, which lends
further support to this picture.

\subsection{Dynamical Structure of the Gas Reservoir}

\begin{figure}
\epsscale{1.15}
\plotone{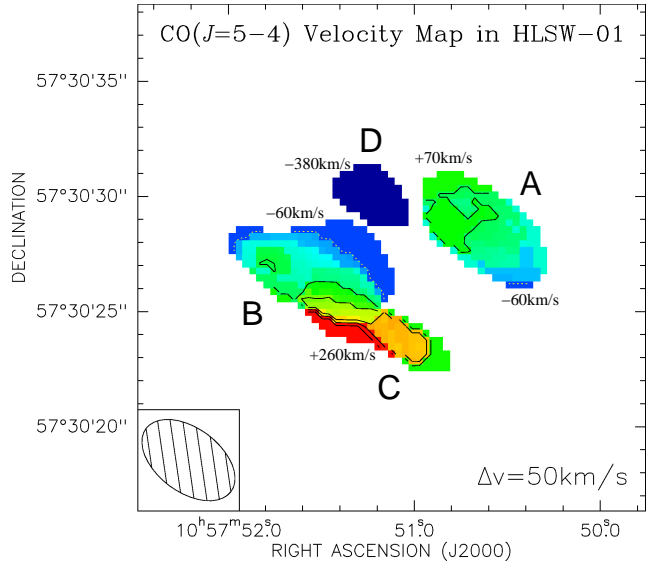}

\caption{\eco\ first moment map. The colors indicate the velocity gradient. 
Contours are shown in steps of 50\,\kms, with dotted (solid) contours
showing blueshifted (redshifted) emission relative to zero velocity in
Fig.~\ref{f2}. `A' to `D' label the four lensed images as in
Fig.~\ref{f3}. Images A and B are blueshifted relative to image C and
redshifted relative to image D. Image A shows a north-south velocity
gradient (red to blue), and image B shows a gradient in the opposite
direction. No velocity gradients are resolved toward images C and D.
\label{f5}}
%
\end{figure}

A first moment map of the \eco\ emission is shown in Fig.~\ref{f5}
(clipped below 2$\sigma$ in each velocity channel, smoothed before
detection). There is a clear velocity gradient between images A/B and
C/D. The peak of image D is blueshifted by up to $\sim$380\,\kms\
(relative to zero velocity in Fig.~\ref{f2}). The peak of image C is
blended with image B, but is consistent with being redshifted by up to
$\sim$150\,\kms.  Images A and B peak on the line center. Image A
shows a smooth north-south velocity gradient from red to blue, and
image B shows a gradient in the opposite direction. This gradient is
also seen in \cco\ emission, indicating that these are likely
dynamically resolved lensed images of the same region in the
galaxy. Images C and D are consistent with being unresolved, (in their
peaks) kinematically distinct regions in the same galaxy or a
companion system. The complexity of the velocity structure is unlikely
to be due to distortion of the velocity field by gravitational lensing
alone, but would be consistent with the dynamical structure of a
major, `wet' (i.e., gas-rich) merger in progress (e.g., Tacconi et
al.\ \citeyear{tac08}; Engel et al.\ \citeyear{eng10}).
 
Based on the velocity-integrated \eco\ map shown in Fig.~\ref{f1}, G11
find a half flux radius of 1.1$\pm$0.5\,kpc for the CO reservoir in
HLSW-01, after correcting for gravitational magnification. However,
the structure seen in the CO velocity map indicates that this size
estimate may be a lower limit. Thus, different parts of the CO
emission may be differentially magnified. However, the lens model by
G11 suggests that differential lensing effects are only moderate
($\lesssim$25\%), even out to a radius of 10\,kpc (their Figure 3),
and thus, have a relatively minor effect on the integrated gas
properties.

\subsection{Physical Properties of the Gas, Dust, and Stellar Components}

The minimum time for which the starburst in HLSW-01 can be maintained
at its current rate is given by the gas depletion
timescale\footnote{An index `0.8' indicates a linear scaling with
$\alpha_{\rm CO}$.}  $\tau_{\rm dep}^{0.8}$=$M_{\rm
gas}$/SFR$\sim$14$\pm$2\,Myr, which is short but within the range of
values found for `typical' SMGs (e.g., Greve et al.\
\citeyear{gre05}).

The ratio between $L_{\rm IR}$ ($\propto$SFR) and $L'_{\rm CO}$
($\propto$$M_{\rm gas}$) can be used as a measure of the star
formation efficiency.  Using $L_{\rm
IR}$=1.43$\pm$0.09$\times$10$^{13}\,(\mu_{\rm L}/10.9)^{-1}$\,\lsol\
(C11), we find a ratio of $\sim$340$\pm$40, comparable to what is
found in `typical' SMGs (e.g., Tacconi et al.\ \citeyear{tac06};
Riechers et al.\ \citeyear{rie10}).

HLSW-01 has a dust mass in the range of $M_{\rm
dust}$=1.0--5.2$\times$10$^{8}$\,($\mu_{\rm L}$/10.9)$^{-1}$\,\msol\
(C11; S11). This corresponds to a gas-to-dust mass ratio of
$f_{gd}^{0.8}$=$M_{\rm gas}$/$M_{\rm dust}$=60--330, and is consistent
with the range of values found in other SMGs (e.g., Michalowski et
al.\ \citeyear{mic10}).

HLSW-01 has a stellar mass of
$M_\star$=6.3$\pm$3.4$\times$10$^{10}\,(\mu_{\rm
L}/10.9)^{-1}$\,\msol\ (C11), yielding a gas mass fraction of $f_{\rm
gas}^{0.8}$=$M_{\rm gas}$/$M_\star$=0.53$\pm$0.29, and a baryonic gas
mass fraction of $f_{\rm bary}^{\rm g,0.8}$=$M_{\rm gas}$/($M_{\rm
gas}$+$M_\star$)=0.35$\pm$0.13. This also yields a specific star
formation rate (SSFR, i.e., SFR/$M_\star$) of 39$\pm$21\,Gyr$^{-1}$,
and a stellar mass doubling timescale ($M_\star$/SFR) of
26$\pm$14\,Myr. These values are consistent with what is found for
other SMGs as well (e.g., Daddi et al.\ \citeyear{dad09a}; Tacconi et
al.\ \citeyear{tac06}, \citeyear{tac08}).

\section{Discussion and Conclusions}

We have detected spatially resolved, highly excited \cco\ and \eco\
emission and \aco\ emission toward HLSW-01, the 250\,$\mu$m-brightest
SMG found in the HerMES {\em Herschel}/SPIRE SDP data.  Based on these
observations, we have refined the redshift of HLSW-01 to
$z$=2.9574$\pm$0.0001.  Once corrected for lensing magnification,
HLSW-01 has a relatively modest molecular gas mass for an SMG. The
$M_{\rm gas}$ in HLSW-01 is about twice that found for the strongly
lensed $z$=2.3 SMG J2135--0102 ($M_{\rm
gas}$=1.4$\times$10$^{10}$\,($\alpha_{\rm CO}$/0.8)\,($\mu_{\rm
L}$/32.5)$^{-1}$\,\msol; J2135--0102 has $\sim$0.87\,times the
apparent 250\,$\mu$m flux of HLSW-01; Swinbank et al.\
\citeyear{swi10}; Danielson et al.\ \citeyear{dan10}; Ivison et al.\
\citeyear{ivi10b}). However, HLSW-01 has a higher CO excitation in its
low- to mid-$J$ lines ($r_{31}$=0.68$\pm$0.03, $r_{53}$=0.51$\pm$0.02,
and $r_{51}$=0.35$\pm$0.02 in J2135--0102; Danielson et al.\
\citeyear{dan10}). This difference becomes even more dramatic in the
high-$J$ CO lines (S11). HLSW-01 also has a higher characteristic dust
temperature of 88$\pm$3\,K (J2135--0102:\ 30--60\,K; Swinbank et al.\
\citeyear{swi10}). The higher CO excitation and dust temperature in
HLSW-01 relative to J2135--0102 and other `typical' $z$$>$2 SMGs may
be due to a higher relative importance of heating by an active
galactic nucleus (AGN), but more detailed studies are required to
constrain the presence of an AGN in HLSW-01. The CO line widths, gas
depletion timescale, star formation efficiency, (limits on) dense gas
fractions, gas-to-dust ratio, gas mass fraction, SSFR, and stellar
mass doubling timescale observed toward HLSW-01 are consistent with
`typical' SMGs (e.g., Greve et al.\
\citeyear{gre05}; Gao et al.\ \citeyear{gao07}). The spatial and
dynamical structure of the molecular gas reservoir is complex,
consistent with what is observed in major mergers in other SMGs (e.g.,
Tacconi et al.\ \citeyear{tac08}; Carilli et al.\ \citeyear{car10};
Engel et al.\ \citeyear{eng10}; Bothwell et al.\ \citeyear{bot10}),
but higher spatial resolution and a full dynamical lens inversion are
required to investigate the velocity structure in more detail (see,
e.g., Riechers et al.\ \citeyear{rie08}). The brightest peaks of the
CO emission show a small spatial offset from those observed in the
rest-frame optical, suggesting that the most gas-rich, most intensely
star-forming regions in HLSW-01 are heavily obscured, consistent with
what is observed in unlensed SMGs (e.g., Tacconi et al.\
\citeyear{tac08}; Daddi et al.\ \citeyear{dad09a}, \citeyear{dad09b};
Carilli et al.\ \citeyear{car10}; Riechers et al.\ \citeyear{rie10}).

HLSW-01 is an exceptional example for the level of detail to which the
molecular gas properties of `typical' SMGs can be studied with the aid
of strong gravitational lensing. Such studies provide crucial insight
into the physical processes that accompany the early stellar mass
buildup during the most active phases in the evolution of massive
galaxies.

\acknowledgments 
We thank the referee for a detailed and helpful report.  DR
acknowledges support from NASA through Hubble Fellowship grant
HST-HF-51235.01 awarded by STScI, operated by AURA for NASA, under
contract NAS 5-26555. AJB acknowledges support from NSF grant
AST-0708653 to Rutgers University. IRAM is supported by INSU/CNRS
(France), MPG (Germany) and IGN (Spain). Support for CARMA
construction was derived from the Moore and Norris Foundations, the
Associates of Caltech, the states of California, Illinois, and
Maryland, and the NSF. Ongoing CARMA development and operations are
supported by the NSF under a cooperative agreement, and by the CARMA
partner universities. The NRAO is a facility of the NSF operated under
cooperative agreement by AUI.

\end{document}